# Linear and Nonlinear Viscoelasticity of Concentrated Thermoresponsive Microgel Suspensions

Gaurav Chaudhary[1,4,7+], Ashesh Ghosh[2,4,8+], Jin Gu Kang[3,4], Paul V. Braun[1-5], Randy H. Ewoldt[1,4,5*] and Kenneth S. Schweizer[2-6*]

[1] Department of Mechanical Science and Engineering, University of Illinois at Urbana-Champaign, Urbana, IL, 61801, USA
[2] Department of Chemistry, University of Illinois at Urbana-Champaign, Urbana, IL, 61801, USA
[3] Department of Materials Science and Engineering, University of Illinois at Urbana-Champaign, Urbana, IL, 61801, USA
[4] Materials Research Laboratory, University of Illinois at Urbana-Champaign, Urbana, IL, 61801, USA
[5] Beckman Institute for Advanced Science and Technology, University of Illinois at Urbana-Champaign, Urbana, IL, 61801, USA
[6] Department of Chemical & Biomolecular Engineering, University of Illinois at Urbana-Champaign, Urbana, IL, 61801, USA
[7] Present Address: School of Engineering and Applied Sciences, Harvard University, Cambridge, MA 02138, USA
[8] Present Address: Department of Chemical Engineering, Stanford University, CA 94305, USA

+ these authors contributed equally to this work

*ewoldt@illinois.edu  *kschweiz@illinois.edu

## Abstract

This is an integrated experimental and theoretical study of the dynamics and rheology of self-crosslinked, slightly charged, temperature responsive soft Poly(N-isopropylacrylamide) (pNIPAM) microgels over a wide range of concentration and temperature spanning the sharp change in particle size and intermolecular interactions across the lower critical solution temperature (LCST). Dramatic, non-monotonic changes in viscoelasticity are observed with temperature, with distinctive concentration dependences in the dense fluid, glassy, and soft-jammed states. Motivated by our experimental observations, we formulate a minimalistic model for the size dependence of a single microgel particle and the change of interparticle interaction from purely repulsive to attractive upon heating. Using microscopic equilibrium and time-dependent statistical mechanical theories, theoretical predictions are quantitatively compared with experimental measurements of the shear modulus. Good agreement is found for the nonmonotonic temperature behavior that originates as a consequence of the competition between reduced microgel packing fraction and increasing interparticle attractions. Testable predictions are made for nonlinear rheological properties such as the yield stress and strain. To the best of our knowledge, this is the first attempt to quantitatively understand in a unified manner the viscoelasticity of dense, temperature-responsive microgel suspensions spanning a wide range of temperatures and concentrations.

## I. Introduction

Colloidal suspensions exhibit rich dynamics that are strongly dependent on externally controllable parameters such as particle packing or volume fraction, the interparticle interaction potential, applied deformation, pH, and temperature [1]–[6]. A wide range of colloidal systems have been explored in the literature, and jamming phase diagrams at ultra-high concentrations have been proposed. While for purely repulsive interparticle potentials the crowding (steric caging) of particles leads to the formation of quiescent colloidal glasses which do not relax structurally on the observational timescales, a fundamentally different dynamic state emerges in attractive colloidal suspensions where particles aggregate, percolate, and kinetically solidify due to strong short range physical bonding [7], [8]. Additional complications arise for soft and deformable microgels [9]–[11] given their internal structure is crosslinked polymer chains imbibed by solvent. This allows such particles to change their volume (and hence the effective suspension packing fraction) and degree of interpenetration depending on factors such as temperature, polymer-solvent interactions, microgel concentration (loading), and osmotic pressure [9], [10], [12], [13].

Poly(N-isopropylacrylamide) (pNIPAM) microgels have been widely employed to study the dynamics of thermosensitive soft colloid suspensions [9], [14]–[18], [53]. Their temperature/pH-sensitivity allows for the *in-situ* control of particle size [54], interparticle-pair potential, and effective packing fraction, all of which affect the structure and dynamics of the suspension. In recent work [9] we studied primarily the low-temperature behavior of such systems where the particles are highly swollen and interact via purely short-range repulsive interactions [13], [19].

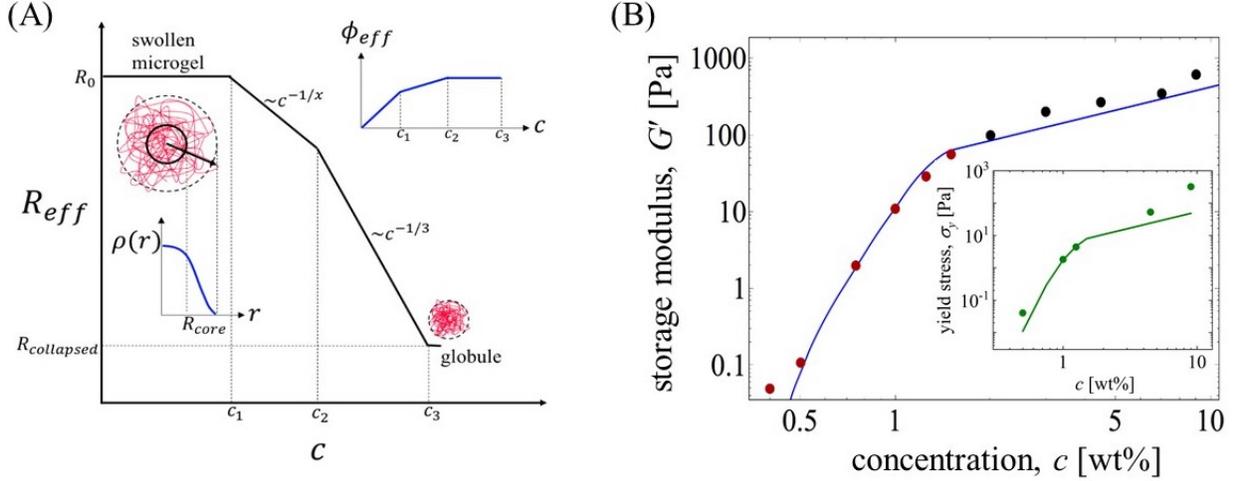

Figure 1 (A) Our physical model of how microgel size and effective volume fraction change as a function of suspension concentration at low temperature where particles repel. (B) Comparison of the experimental (points) and theoretical (lines) linear elastic shear modulus and yield stress (inset). Red data points correspond to the concentrations below 'soft jamming' crossover, and black points correspond to the 'soft-jammed' regimes. Data for the figure are adapted from our prior work [9].

Others have also studied their transition from a repulsive glass-like to attractive gel-like state upon heating a highly packed pNIPAM microgel suspension beyond the LCST [15], [20], [21], [55-56].

In our recent joint experimental-theoretical work [9] we focused on understanding the linear and nonlinear rheological response of purely repulsive slightly charged pNIPAM microgel suspensions at low temperature (well below the LCST) [9] over a very wide range of concentrations spanning the dense Brownian fluid, glassy fluid, and 'soft-jammed' regimes. In the glassy fluid regime, experiments found the linear elastic shear modulus grows over 3 orders of magnitude with concentration following an apparent power law with an exponent of ~ 5.6. Ultimately a crossover to linear concentration dependence occurs which signals the onset of "soft jamming". Representative central findings germane to our present article are illustrated in Fig. 1. The precise modeling of the size of soft microgels with a core-corona microstructure is challenging. Fig. 1A schematically describes the qualitative effect of increasing microgel concentration on particle size at a fixed temperature well below the LCST. We built on recent experiments [22], [23] focused on understanding the concentration-dependence of microgel particle size at low temperatures to construct a model for the microgel effective radius ($R_g$) which has distinct power-law concentration-dependences above ($R_g \sim c^{-1/3}$) and below (much weaker dependence $R_g \sim c^{-1/6}$) a critical concentration ($c_2$) identified as the 'soft-jamming' crossover. To note, at very low concentrations ($c < c_1$), particles have a fixed size consistent with the dilute swollen microgel state.

Using a suite of microscopic liquid state statistical mechanical theories for the structure and dynamics of microgel particles modeled as soft repulsive Hertzian spheres with the size dependence per Fig. 1A, we made predictions for the linear storage modulus and yield stress that were compared with our experiments (Fig. 1B). Good qualitative and quantitative agreements were obtained, including the $G' \sim c^{5.6}$ (or, $G' \sim c^1$) dependences below (above) $c = c_2$; additional theoretical predictions were made for equilibrium pair structure of soft microgels and structural relaxation time. In the linear viscoelastic measurement, the experimental finding of concentration dependent power-law regimes with exponents of ~5.6 (for $c < c_2$) and ~1 (for $c > c_2$) appears distinct from prior efforts of ionic microgels [13]. Trends in the yield stress of the suspensions also compared well with the theoretical calculations.

In this article we build on our recent progress to explore with integrated experiment and theory the role of temperature and concentration on the viscoelastic properties of the same slightly charged pNIPAM microgel suspensions. We first experimentally study the dynamical behavior over a wide range of concentration and temperature. Varying temperature has two salient effects: (1) microgel particles continuously de-swell as a function of temperature below the LCST, followed by a sharp decrease in the size above the LCST, and (2) the interparticle-pair potential changes from purely repulsive at low temperature to one that has an additional strong short-range attraction above the LCST at higher temperature. We extend and apply our theoretical model [9] to understand the temperature-dependent dynamics with a primary focus on the change in the linear shear elastic modulus below the LCST, and its dependence on concentration at temperatures well above the LCST.

Our experimental methods and results are described in sections II and III, respectively. A brief review of the basics of the well-established statistical mechanical theory methods we employ is provided in section IV. The theoretical methods are then extended and applied to study a specific model of microgels below and above the LCST in section V, and the results are compared with our experiments. The article concludes with a discussion and summary of key findings. Additional experimental details and results can be found in the supporting information (SI).

## II. Materials and Methods

### A. Microgel synthesis and characterization

Slightly charged self-crosslinked pNIPAM microgels were synthesized under a 'cross-linker free' condition following the protocol described in literature [24] (and our prior article [9]) with a few modifications. Briefly, free-radical polymerization of NIPAM in water was initiated using potassium persulfate in the absence of added cross-linker. Stable nanospheres, not linear chains, are formed if the solution is incubated at temperatures well above the LCST of PNIPAM (~32°C). Self-crosslinking by chain transfer reactions during and after polymerization lead to the formation of gel nanospheres [25]. Although microgels prepared with a similar preparation protocol have been referred to as charge neutral [24], [26], we refer to them as 'slightly charged' because the initiator used may possibly leave some charge on the colloids. A stock solution of $c = 9\ wt\%$ was then diluted with deionized water to achieve the desired concentration of the slightly charged microgel suspension. The particle radius was determined by dynamic light scattering (DLS) (Zetasizer Nano ZS, Malvern) and a Helium-Neon gas laser emitting at $632.8\ nm$ on a very dilute suspension ($0.04\ wt\%$) with a beam diameter of $0.63\ mm$.

### B. Rheological Characterization

Rheological experiments are performed on a torque-controlled rotational rheometer (model Discovery Hybrid 3 from TA instruments and model MCR702 from Anton Paar) with plate-plate geometry. To suppress wall slip, a 600 grit, adhesive-back sandpaper (Norton Abrasives) was adhered to the contact surfaces. Plate diameters were chosen to obtain a measurable torque response higher than the minimum torque resolution [27]. A 60 mm plate was used for dilute samples of 0.03−0.25 wt%, a 40 mm plate for 0.25−1.5 wt%, a 20 mm plate for 0.5−4.5 wt%, and a 8 plate mm for 4.5–9 wt% samples. The typical gap in all experiments was between 550–750 μm, far larger than the particle size, thus eliminating confinement effects. A solvent trap, with a wet tissue adhered to its interior, was used to minimize solvent evaporation during the measurements. The temperature of the bottom plate was controlled using a Peltier system. Samples were rejuvenated by shearing at 50 s$^{-1}$ for 60 s and then allowed to relax for 120 s before taking measurements to suppress aging effects and erase any history [19].

Two types of rheological characterization were performed: oscillatory shear and steady shear. To probe the linear response at different fixed temperatures, frequency sweeps were performed from $\omega = (0.03-100)$ rad/s at a strain amplitude of 1%. Temperature ramps were performed from 10-40°C at fixed angular frequency of 1 rad/s, and a strain amplitude of 1%. Steady shear experiments were also performed with shear rates varying from $(300 - 0.01)$ s$^{-1}$ while waiting for the system to reach an apparent steady state as deduced by < 5% variation in torque over a period of 30 s.

## II. Experimental Results

### A. Temperature Dependent Particle Volume

The microgel particles have a mean diameter of ~550 nm at 10°C at a dilute concentration (0.04 wt%). The dependence of their hydrodynamic diameter on temperature at low concentrations ($c = 0.04$ wt%) is shown in Fig. 2A. In both heating and cooling cycles, we observe a sharp change in microgel size across the LCST (~ 32°C). The swollen particle collapses to a dense globular state of mean diameter of ~ 250 nm at 45°C. Such a large variation in size is attributed to the "intramolecular" phase transition of pNIPAM polymer chains at the LCST due to the change of interaction between the solvent and polymer from good to poor. As a result, the volume of solvent a 'free' microgel can imbibe in the equilibrium configuration changes drastically [28]. A modest hysteresis in the hydrodynamic diameter while cooling and heating the suspension is observed, consistent with prior studies [12], [29]. Microgels prepared by such a crosslinker free preparation protocol are known to be highly deformable and ultra-soft [24]. Our prior estimates for the individual particle modulus is ~ 1.5 kPa in the fully swollen state [9], [12].

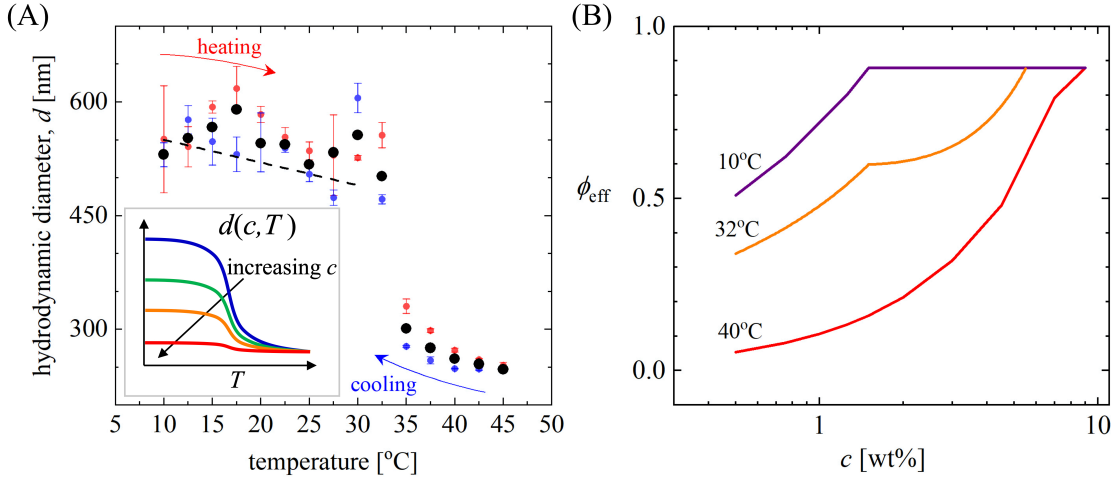

Figure 2 (A) Temperature dependence of the hydrodynamic diameter of pNIPAM microgel particle measured using dynamic light scattering of a dilute aqueous suspension of microgels. The points in red correspond to a temperature increase cycle and the blue points correspond to a temperature decrease cycle. The black data points are the average of heating and cooling cycles. The "intramolecular" phase transition of the pNIPAM polymer across the LCST leads to a drastic change in the particle size. (inset) Schematic showing the expected response at higher concentrations. (B) The estimated microgel suspension volume fraction at various concentrations and temperatures; details of the estimation are provided in section IV A.

A schematic of the expected temperature dependence at higher concentration is shown in the inset of Fig. 2A. With increasing concentration, we expect the low temperature microgel radius follows Fig. 1A due to modest steric deswelling induced by osmotic (compressive) steric particle-particle interactions. Estimates of the microgel effective packing or volume fraction at T=10°C were obtained per our prior work [9]. Here we provide estimates of how the temperature affects the effective packing fraction via change of effective radius. The precise theoretical modeling of the effective packing fraction is described in later sections. However, the crucial qualitative point is that for any fixed concentration, with increasing temperature (equivalently with decreasing radius) the effective packing fraction decreases as shown in Fig, 2B. The degree to which this occurs depends on the specific concentration range studied -- fluid, glassy, and the "soft-jammed" regimes, which exhibit different functional dependences of $R$ on $c$ (Fig. 1A).

### B. Temperature Dependent Linear Rheology

At low temperature (10°C), the microgel suspensions for $c > 0.4$ wt% exhibit a predominantly solid-like response since the storage modulus $G'$ is larger than the loss modulus $G''$ over the entire probed frequency range. Moreover, $G'$ is nearly frequency-independent (SI Fig. S1), indicating the structural relaxation time is very long and the system does not flow on the experimental time scale. We also found that the response at different temperatures (above and below the LCST) is nearly frequency-independent for a subset of concentrations (SI Fig S2). The linear moduli ($G'$, $G''$) at a fixed strain amplitude $\gamma_0 = 1$ % as a function of the temperature are shown in Fig. 3A-B over a wide concentration range (0.4-9 wt%). We note that the $c = 0.4$ wt% sample gives a very weak torque response at all probed temperatures which can be clearly observed as the noise in the sinusoidal response of the material [27]. Above this concentration, the torque signals are significantly larger than the manufacturer-specified low torque limit (below the LCST) and hence can be reliably resolved.

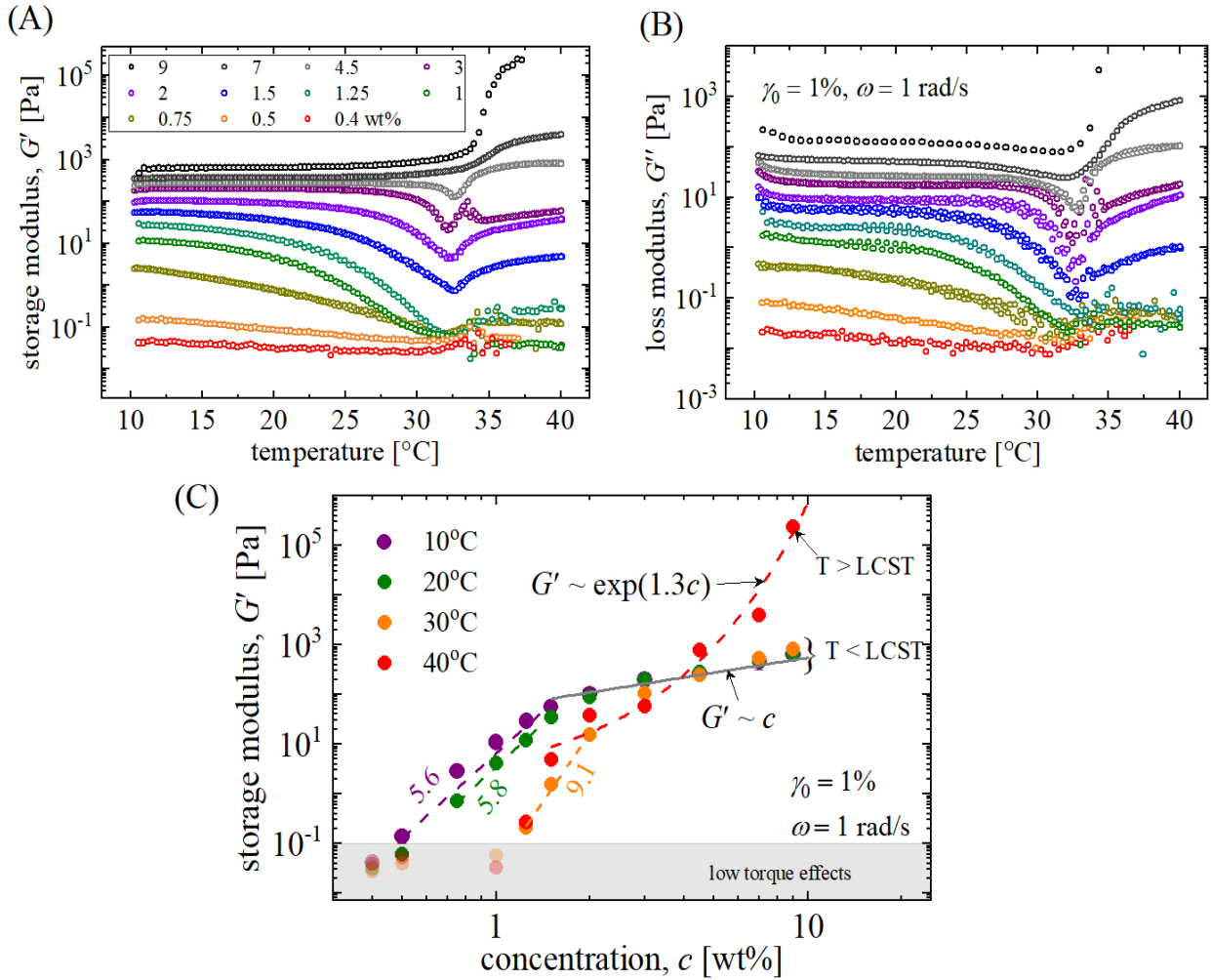

Figure 3 Temperature and concentration dependences of the linear shear rheology. (A) Storage modulus, G′ and (B) Loss modulus G″ for various concentrations of pNIPAM microgel suspensions as a function of temperature at an angular frequency of 1 rad/s. The suspensions show a rich dependence on the temperature. (C) Concentration dependence of linear elastic shear modulus, G′, at various temperatures. At temperatures below the LCST, i.e. 10, 20 and 30°C, G′ exhibits two regimes of concentration dependence: a strong apparent power-law dependence (G′ ~ $c^n$, n > 1) followed by a roughly linear dependence. Above the LCST (40°C), the modulus monotonically increases with the concentration due to increasing strength of attractive interactions between the microgel particles under poor solvent conditions.

The diverse rheological response of the suspensions at various concentrations to changing temperature is seen from Fig.3A-B as the system is heated up to and beyond the LCST. Experimentally, visible changes in the transparency and color of the suspension are observed. Local microgel aggregation is also observed at low concentrations as the attractive intra- and interparticle interactions emerge above the LCST (SI Fig. S3). Over the entire concentration regime, the temperature-dependence of the linear viscoelastic response can be broadly divided into three categories: weak softening, strong softening-to-stiffening, and stiffening, corresponding to a dramatic *non*-monotonic evolution with heating.

In the lower concentration range (0.4-1.25 wt%), the moduli monotonically decrease as the temperature is raised from 10 to 30ºC (below the LCST). From our prior work [9], we concluded that the suspension in this concentration regime at 10ºC behaves as a glassy liquid with purely repulsive interparticle interactions. In a physical sense, each microgel particle is trapped in a cage formed by its many neighbors. Such cages store elastic energy under small deformations, providing a dominant elastic suspension response [7]. The time scale of collective relaxation of such cages under small deformations is generally much larger than the experimental time scales, and hence the suspension behaves as a solid ($G' > G''$) on the observational time scales. From 10 to 20ºC, similar physics appears to determine the overall dynamics. A weak softening in the shear moduli can be observed, likely due to a decreasing effective volume fraction as particles mildly de-swell in this temperature regime (Fig. 2A). A rather sharp decrease in the values of moduli is observed between 20 to 30ºC at concentrations of 0.75, 1, and 1.25 wt%; in contrast, for the concentrations 0.4 and 0.5 wt% less change is observed. A further increase in temperature from 30 to 40ºC leads to dramatic non-monotonic trends of the shear moduli with temperature and concentration. These non-monotonic trends can be attributed to the spatially heterogeneous clustering of de-swollen microgel particles, which do not form a percolated network (evident in SI Fig. S3), as the attractive interactions between the particles emerge across the LCST.

The suspensions in the higher concentration regime (1.5-4.5 wt%) at low temperatures (10-20ºC) show a similar weak softening response as observed when $c < 1.5$ wt%. The softening effect weakens with the increasing concentration in this regime. Increasing the temperature from 20-32ºC leads to drastic softening of the linear moduli, but the suspension maintains a predominant elastic response ($G' > G''$). A further increase in temperature above the LCST, i.e. in the range 32-40ºC, results in a monotonic increase of the linear moduli. We hypothesize such a softening-to-stiffening trend in the linear moduli is a consequence of two competing effects which are *both* temperature and absolute concentration dependent: (1) reduction of effective volume fraction of particles across the LCST, and (2) the emergence of attractive interactions between the particles above the LCST. Our expectation is that well above the LCST, the effective volume fraction of a suspension decreases weakly (since the DLS data in Fig. 2A suggests fairly weak dependence of particle size on temperature above the LCST), and the increase in moduli is directly related to the increasing interparticle attractions and physical bond formation. It is well established that the attraction strength between pNIPAM polymers increases with temperature above the LCST (poorer solvent quality), and the formation of particle networks provides a dominant solid-like response. At the lower concentrations (0.4-1.25 wt%), even though the interactions between the particles switch from repulsive to attractive, we suspect that the particle volume fractions at high temperatures are low enough that no space spanning (percolated) network is present.

Finally, for $c$ = 7 and 9 wt% (soft jamming regime for repulsive particles), the linear moduli do *not* show any apparent softening below the LCST, yet another qualitatively different behavior. Rather, the moduli exhibit a very weak stiffening response as temperature is increased from 10-32ºC. Further heating leads to a drastic stiffening of the moduli, which is more prominent at 9 wt% compared to 7 wt%. The trends above the LCST can be attributed to the emergence of strong attractive interactions between particles in *highly crowded* suspension which experience dynamical constraints associated with both steric caging and strong physical bonding. Below the LCST, the weak increase in the moduli is suggestive of the weak dependence of the effective interaction potential on the temperature [30]. At such high concentrations, the microgels are highly compressed close to their collapse size due to steric crowding effects [9], [31], hence we expect temperature may not have as drastic effect on the effective volume fraction, contrary to the lower concentration regimes. These physical ideas will be quantitatively explored in sections IV A-B.

In earlier work by others on PNIPAM suspensions [18], it was shown that at an intermediate temperature between the repulsive glassy and attractive particle-gel state, a liquid-like state is achieved (also called re-entrant glass melting regime). However, in the concentration range of 1.5-9 wt%, we do not observe such behavior, and the samples remain predominantly elastic, $G' > G''$, over the entire temperature range. Below 1.5 wt% it is not clear from the experimental data whether such a liquid state is realized due to instrument low torque limitations. Another difference that we observe is in the frequency dependence of linear viscoelastic moduli above the LCST. Previous work by others [19], [20] found a weak power-law dependent viscoelastic modulus. We have not explored the frequency dependence systematically at all concentrations and temperatures, but our frequency-dependent experiments (Fig S2) at selected concentrations and temperatures do not establish a clear power-law dependence. Several experimental limitations render the measurements impossible over a very wide range of frequencies [27].

The linear moduli at all temperatures generically exhibit a monotonic increase with the concentration. Fig. 3C shows the concentration dependence of $G'$ at various temperatures. The measurements at low concentrations (0.4 – 0.5 wt%) at 10 and 20ºC agree reasonably well with the characteristic modulus scale for dense hard-sphere suspensions [17], [32], $G' \sim k_B T/(2R)^3$, where $k_B$ is the Boltzmann constant, $T$ is the temperature, and $R$ is the sphere radius. For our microgel system, such a scaling estimate gives a value of $G = 0.024$ Pa for $2R = 550$ nm. At 10, 20 and 30ºC the elastic modulus shows a dramatic dependence on concentration ($G' \sim c^n$, $n > 1$) in the range (0.4 - 1.25 wt%) followed by a weaker concentration dependence ($G' \sim c$) in the range (1.5 – 9 wt%). Such a trend has also been observed in several other studies on related systems with different power-law exponents, $n$, which depend on various material-specific aspects such as particle stiffness and chemistry [13], [19]. The power-law fits for various temperatures are shown in the Fig.3C. The exponent $n$ increases with increasing temperature below the LCST. This is likely a consequence of increasing particle stiffness as the particle volume decreases with increasing temperature, and is consistent with prior literature [19]. We note that at 30 and 40ºC, the measurements at low concentration are influenced by the low torque limit in the rheometer and the apparent plateau-like response at these temperatures is likely a consequence of such an experimental artifact and is not the true rheological response.

The transition in the power-law response at 10, 20 and 30ºC has been attributed to a 'soft jamming' crossover in repulsive soft glasses in several studies in the literature [13], [19]. The "soft jamming" crossover takes place at a higher concentration as temperature is raised from 10 to 30ºC, as expected. Because of the discrete concentrations chosen in experiments, we cannot precisely establish the "soft jamming" crossover concentration as a function of temperature. However, it is apparent from our data that such a trend persists. This is because of the temperature induced particle de-swelling which results in a higher particle number density (concentration) to reach the effective volume fraction necessary for such a crossover. In our prior work [9], we estimated the transition occurs at an effective volume fraction of ~88%. We also attributed the linear relationship between elastic modulus and concentration above the soft jamming crossover to the strong shrinking of repulsive microgel particles which results in *a constant* effective volume fraction, and hence a modulus that grows linearly with number density of particles as the suspension concentration is increased. At 40ºC, a fundamentally different response is seen where unlike the observations of a soft jamming regime at lower temperatures, we observe a monotonic modulus increase with concentration. Interestingly, the modulus closely follows an *exponential* dependence on concentration at this temperature. Physically we attribute this modulus increase to increasing attractive interactions and the emergence of physical bonds between compact microgel particles at high concentrations.

### C. Nonlinear Rheology

Figures 4 A-B show the steady shear flow curves for suspensions at two representative concentrations (3 and 7 wt%) at various temperatures (see SI Fig. S5 for other concentrations). As seen in the previous section, these concentrations show very different thermoresponsive linear elastic responses as the temperature is raised beyond the LCST. Their nonlinear response resembles that of a yield-stress fluid at all temperatures, with the stress reaching an apparent plateau value at low shear rates. At $c$ = 3 wt%, the flow curves appear to shift vertically downwards at higher temperatures (Fig. 4 A). Such a trend is not observed for $c$ = 7 wt%, with the flow response showing a non-monotonic temperature-dependence (Fig. 4 B). We could not explore the flow response at temperatures above the LCST due to significant evaporation during the duration of the mechanical test, but the high shear rate viscosity at all the concentrations above the LCST was significantly lower than the corresponding values below the LCST.

The flow curve in the studied concentration and temperature range is well captured by the phenomenological Herschel-Bulkley (HB) model [32], [33] given by

$$\sigma(\dot{\gamma}) = \sigma_y^{HB}\left(1 + \left(\frac{\dot{\gamma}}{\dot{\gamma}_c}\right)^n\right) \tag{1}$$

where $\sigma_y^{HB}$ is the apparent yield strength and $n$ is the flow index. The characteristic shear rate, $\dot{\gamma}_c$, is interpreted as a crossover shear rate from rate-independent plastic flow to rate-dependent viscous flow. Equation (1) is fit using variance weighting to each dataset at a given concentration and temperature. For $c$ = 3 wt%, $\sigma_y^{HB}$ decreases with temperature below the LCST, similar to the trends of decreasing $G'$ at this concentration (both $\sigma_y^{HB}$ and $G'$ decrease roughly by a factor of 2 as temperature is increased from 10 to 30 ºC) as shown in Fig. 4A, while $\dot{\gamma}_c$ exhibits a monotonic increase as shown in Fig. 4A inset. The exponent $n$ increases with temperature, varying from 0.43 to 0.56.

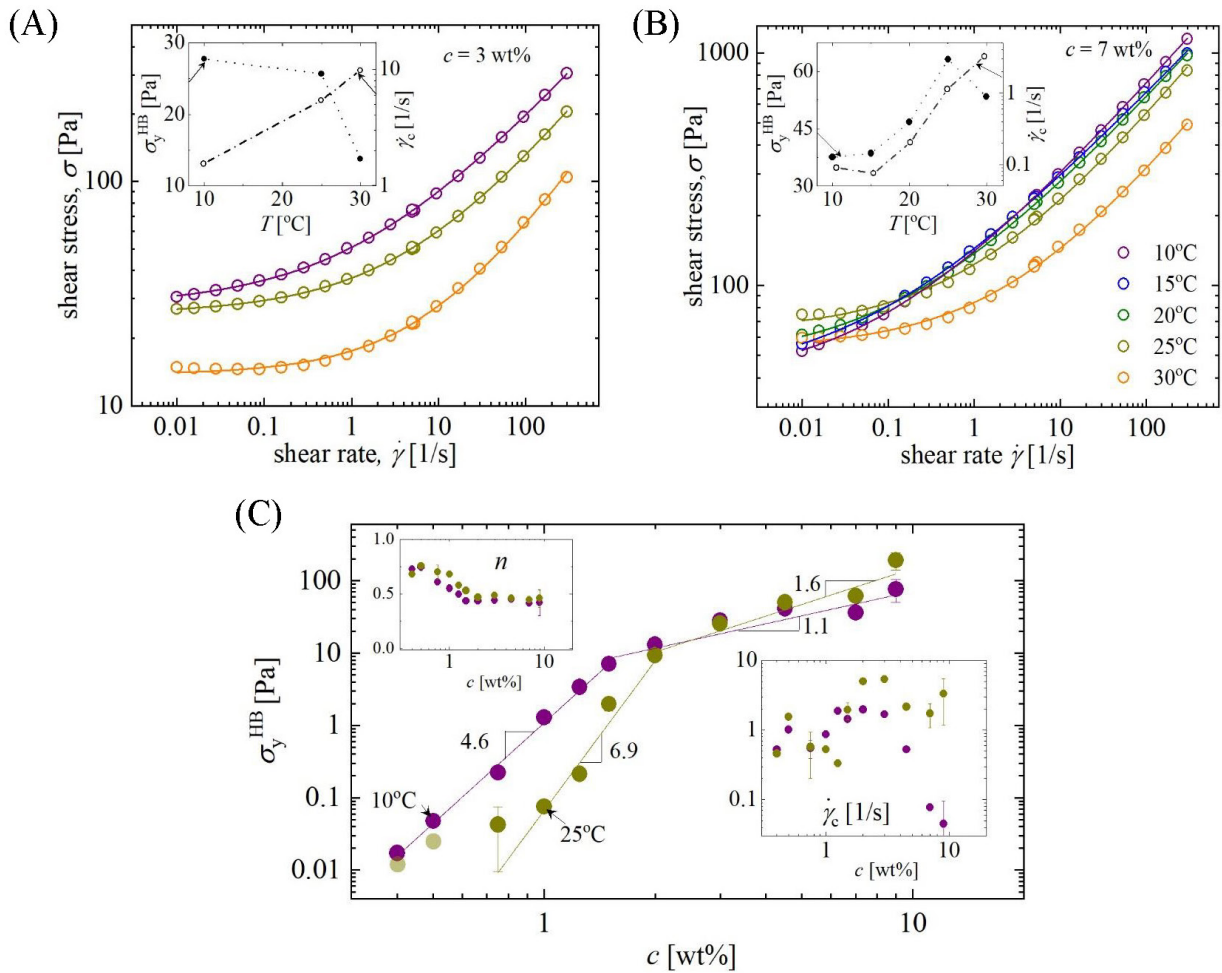

Figure 4. Steady shear flow properties with Herschel-Bulkley (HB) model fits at different temperature and concentration. (A,B) Data for c = 3 wt% and 7 wt%, respectively. Lines are fits with Eq.(1); select fit parameters in inset. (C) Herschel-Bulkley fit parameters for the entire range of concentrations at 10ºC (purple) and 25ºC (yellow). The solid lines are least squares power-law fits.

We attribute such trends to the increase in suspension fluidity at high shear rates and temperatures. For $c = 7$ wt%, such a trend is not followed (Fig. 4B). The apparent static (zero shear rate limit) yield stress, $\sigma_y^{HB}$, initially is nearly independent of temperature, but then increases, reaching a maximum value (at ≈ 25ºC), followed by a weak decrease at the highest probed temperature (30ºC) (both $\sigma_y^{HB}$ and $G'$ marginally increase a factor of 1.4 and 1.2, respectively, as temperature is increased from 10 to 30 ºC). Note that the change of the flow curves, and hence the HB fit parameters, is very subtle with temperature. Similar non-monotonic trends were also observed at $c = 4.5$ and 9 wt% (see SI Fig. S5). We attribute such non-trivial effects in the steady shear response to the complex nature of interparticle interaction potential, and the varying effective volume fraction of the suspension with temperature.

Figure 4C shows the Herschel-Bulkley fit parameters as a function of suspension concentration at two temperatures: 10 and 25°C. Results for the flow curves at all concentrations and corresponding HB fits are given in SI Fig. S4. Similar to the observations made earlier for $G'$ at the temperatures below the LCST in Fig. 3C, $\sigma_y^{HB}$ follows a strong apparent power-law concentration dependence below the "soft jamming" crossover, followed by a weak concentration dependence at both temperatures. Power-law indices in the different concentration regimes are also temperature-dependent. The yield stress parameter $\sigma_y^{HB}$ at 25°C is lower than that at 10°C below a certain concentration, followed by an opposite trend beyond it. Above the "soft jamming" crossover, both the linear elastic shear modulus $G'$ and $\sigma_y^{HB}$ exhibit a nearly linear dependence on concentration at 10°C, while at 25°C the power-law concentration dependence of $\sigma_y^{HB}$ deviates from linearity unlike $G'$. Such differences further exemplify the complex suspension dynamics due to coupled effects from varying interparticle interactions and effective volume fraction with temperature.

SI Fig. S6 shows the comparison of relative viscosity at high shear rate (300 s$^{-1}$) at the two temperatures in the limit of dilute concentrations, and the fit value of intrinsic viscosity exemplifies the temperature-dependent flow dynamics. Differences can also be observed in the trends exhibited by $\dot{\gamma}_c$ as shown in the inset of Fig. 4 C. The characteristic shear rate at 10°C has a nearly constant value below the "soft jamming" transition followed by a decreasing trend above it. At 25°C, no specific trends could be observed, and $\dot{\gamma}_c$ is roughly constant with much larger magnitudes compared to 10°C. The flow index, $n$, shows similar trends at both temperatures, decreasing monotonically with the concentration below the soft jamming transition, followed by a nearly constant value of 0.41 in the soft jammed regime. The values of $n$ are marginally higher at 25°C at all concentrations indicating an increased fluid-like nature of the suspension at high temperature undergoing deformation at large shear rates.

## III. Modeling Microgels at Low and High Temperatures

### A. Low Temperature Repulsive Regime

Simulations and theory have previously modeled the core-shell microgel interparticle interaction as a central repulsive Hertzian contact pair potential [35,36] at low temperature [9], which is given as

$$\beta V(r) = \begin{cases} \dfrac{4E}{15}\left(1 - \dfrac{r}{d}\right)^{\frac{5}{2}} & r < d = 2R_{eff} \\ 0 & r \geq d \end{cases} \quad (2)$$

Here, $\beta = (k_B T)^{-1}$ is the inverse thermal energy, $r$ is the interparticle separation, and $d$ is the particle diameter-like length scale. The front factor $4E/15$ plays the role of an inverse dimensionless temperature which controls the elastic stiffness of a particle. The dimensionless parameter $E$ is proportional to the Hertzian contact stiffness normalized by the temperature, and depends on the particle Young's modulus $Y$, diameter, and Poisson ratio (Equation (3) of our prior work [9]), scaling as $E \sim \beta Y d^3$. Comparison to our prior low temperature experiments provided the estimate of $E = 30,000$ for the pNIPAM microgels studied here [9]. For a microgel composed of cross-linked flexible polymer chains, $Y \sim \rho_x k_B T$ with $\rho_x$ the crosslink number density ($= N_x/d^3$) per synthesis protocol. For a fixed crosslink density, combining the scaling laws results, to a leading order, in a $E$ value independent of the changing particle diameter $d$. As a minimalist assumption, we will use the same $E$ to model the $T$ dependence in this study.

Temperature sensitive pNIPAM microgels show a continuous but abrupt coil (low T) to globule (high T) transition upon traversing the LCST at ~31-33°C [9]. With increasing temperature far enough beyond the LCST, the microgels expel more water and eventually achieve a dense collapsed globule state. Thus, at T>37°C we expect particles have effectively collapsed to an essentially temperature independent size. At low temperature, the microgel diameter changes linearly with heating until very close to LCST, per dynamic light scattering measurements [37]. This initial linear diameter decrease with heating dependence can be modeled as [37],

$$d(c, T) = d(c, T_0) - \alpha(c)(T - T_0) \quad (3)$$

where the diameter also changes with concentration and $T_0$ is a known low temperature value. Existing models of microgel size as a function of temperature [37] can be obtained by setting $\alpha(c)$ to be a concentration independent fixed number. For our purpose, the individual concentration and temperature dependences of microgel size that have been reliably modeled previously [9] is extended to *simultaneously* describe microgel size as a function of both concentration and temperature. We emphasize that temperature dependent de-swelling is different from concentration dependent reduction of microgel size since the latter it is driven by a qualitatively different physical mechanism: inter-microgel steric compression or osmotic forces versus reduction of solvent quality. This is reflected in the second term in Eq. (3) which effectively *decouples* the concentration and temperature dependences as a simple approximation. As the concentration grows, the soft microgels eventually are forced into contact and interpenetrate (soft jamming crossover). In our previous work [9], we provided a prediction of how microgel size at low $T$ changes with growing concentration, and discussed in detail the different concentration dependences below and above the soft jamming crossover.

For our pNIPAM system we find $\alpha = 3$ for dilute suspensions per Eq. (3) (the dashed line through the data for temperatures below the LCST in Fig 2A). As discussed above and previously [9], the concentration dependent steric de-swelling effect places the higher concentration (c>4.5wt%) microgel samples at an initial size that is already collapsed to its dense globule-like state even in the low temperature regime, and hence they cannot expel more water with increasing temperature. Thus, these microgel samples are expected to have reached their maximum internal (melt-like) density limit. From Fig. 3A, we see that for c>4.5wt% the storage modulus is independent of temperature for T=10-32°C. We hypothesize that for these concentrations, steric de-swelling causes the particles to have a similar collapsed globule size, and very small temperature-dependent size changes can be safely ignored (i.e. $d(c,T) = d(T)$= constant for temperatures below the LCST *and* c>$c_{threshold}$). Thus, the effective suspension volume fraction remains constant, implying for low concentrations (c=0.5wt%), $\alpha(c) = 3$, and for c>$c_{threshold}$ one has $\alpha(c) = 0$. Consistent with the storage modulus data in Fig 3A, we assume $\alpha(c)$ goes to 0 at c=5.5wt%. Examination of the storage modulus data suggests this picture is reasonable for intermediate concentrations of 4.5wt% and 7wt% where the ratio $\frac{G'(10°C)}{G'(32°C)}$ goes to 1, taken as a signature of non-changing size (hence constant suspension volume fraction) of the microgels in the temperature range $10 - 32°C$. Thus, as a *simple* model guided by our experimental observations, we assume $\alpha(c)$ decays linearly from 3 to 0 with increasing concentration: $\alpha(c) = \alpha_0 - 0.60 \times (c - 0.5)$, where $\alpha(c = 0.5wt\%) = \alpha_0 = 3$. Of course, the precise concentration where $\alpha(c)$ goes to 0 could be taken anywhere in the narrow range 5-6wt%, but we have found that our theoretical results presented below are not very sensitive to this choice, which we emphasize is used only for the 5.5wt% sample.

To summarize, the microgel diameter changes with temperature per Eq. (3) where $d(c, T_0) = d(c, 10°C)$ is set by the models adopted a low temperature before significant size change begins. Thus, below the LCST temperature we have

$$\phi(c; T) = \phi(c; T = 10°C) \times \left(\frac{d(c; T)}{d(c; 10°C)}\right)^3. \tag{4}$$

A plot of this volume fraction at 32°C is shown in Fig. 2B. The conversion factor for storage modulus from dimensionless units ($k_B T/d^3$) to Pascals also follows a similar relation due to changing size and volume fraction, where with increasing $T$, this dimensionless unit increases as $d$ decreases.

**B. High Temperature Attractive Regime**

Above the LCST the interactions between microgels become attractive due to strong hydrophobic attractions [38-39] that emerge in a rather abrupt, but smooth, fashion. A model effective pair potential that applies for the entire range of temperature of interest (10-45°C) where the complex microgel swelling de-swelling behavior takes place does not exist. We thus explore the following minimalist, 2-parameter model that contains an attractive pair potential,

$$\beta V(r) = \begin{cases} \beta\epsilon + \frac{4E}{15}\left(1 - \frac{r}{d}\right)^{\frac{5}{2}}, & r < d \\ \beta\epsilon e^{-\frac{r-d}{a}}, & r \geq d \end{cases} \tag{5}$$

where $a$ is the range of the attractive interaction and $\epsilon < 0$ is its strength. The hydrophobic force law is known [40,41] to follow an exponential decay with distance between two large micron sized spheres (essentially two surfaces), with a small mean decay length related to a water molecule diameter ($d_{H_2O} = 0.3\ nm$). This motivates our fixing of $a = 0.01d$, where $d = 250\ nm$, corresponding to a reasonable range of a few nanometers. Thus, only one unknown parameter remains: the strength of the hydrophobic attraction when a pair of microgels at "contact" (r=d), which based on the known physical chemistry of the hydrophobic effect [40-42] is expected to be temperature dependent.

Finally, since we do not know how to precisely turn on the attraction mathematically as a continuous function of temperature, we focus only on understanding the experimental behavior at the highest temperature studied in experiments where inter-particle attractions are strongest.

**C. Packing Structure, Dynamic Localization, Elastic Modulus, and Yielding**

Given the interparticle pair potential, we can theoretically determine the equilibrium packing structure and use it to predict the equilibrium Brownian dynamics, activated relaxation, and rheological properties. Since all details of the theory and methods used here have been exhaustively described in prior papers [9, 43, 43-47], we only summarize the key aspects needed for our present work which treats microgels as soft particles at the center-of-mass level.

The interparticle pair correlation function, $g(r)$, static structure factor, $S(k)$, and, direct correlation function $c(r)$ can be obtained from Ornstein-Zernike (OZ) integral equation theory [48] that relates the non-random part of interparticle pair correlation function $h(r) = g(r) - 1$ and the direct correlation function $c(r)$ via,

$$h(r) = c(r) + \rho \int c(|\vec{r} - \vec{r'}|)h(r')d\vec{r'} \tag{6}$$

where $\rho$ is the particle number density. Collective density fluctuations are quantified in Fourier space (wavevector, k) by $S(k) = 1 + \rho h(k) = (1 - \rho c(k))^{-1}$. Numerical solution of Eq. (6) requires a closure approximation that connects the pair correlation functions with the interparticle potential ($\beta V(r)$) and thermodynamic state (density, temperature) of the system. For soft colloids, we employ the well-studied hypernetted chain closure (HNC) [48],

$$c(r) = -\beta V(r) - \ln(g(r)) + h(r) \tag{7}$$

Sample calculations of S(k) are shown in Fig. 5A that illustrate how this key structural quantity evolves with increasing attractive interactions.

The above structural information then enters as the essential input to the naïve mode coupling theory (NMCT) description of ideal dynamical arrest formulated in terms of effective forces quantified by pair correlations [49-50]. The self-consistent single particle dynamic localization length equation is given by [48,50]

$$\frac{1}{r_L^2} = \frac{\rho}{18\pi^2} \int_0^\infty dk \, k^4 C(k)^2 S(k) e^{-\frac{k^2 r_L^2}{6}\left(1+S^{-1}(k)\right)} \quad (8)$$

where $r_L$ is the dynamic localization length. The existence of a non-infinite solution beyond a threshold volume fraction, $\phi_c$, signals the onset of a transiently localized or solid-like regime and a crossover to an activated barrier hopping controlled dynamics and mechanics. One can derive a (high-frequency) dynamic elastic modulus formula starting from a stress-stress correlation function as [51]

$$G' = \frac{k_B T}{60\pi^2} \int_0^\infty dk \left[ k^2 \frac{d}{dk} \ln(S(k)) \right]^2 e^{-\frac{k^2 r_L^2}{3S(k)}} \approx a\phi \frac{k_B T}{d r_L^2} \quad (9)$$

where $a$ is a numerical pre-factor, and the elastic shear modulus varies inversely with localization length.

Beyond the ideal NMCT localization "transition" (a crossover), activated dynamics is treated based on the nonlinear Langevin equation (NLE) theory formulated in terms of dynamic free energy, $F_{dyn}(r)$. The latter controls the force on a tagged particle due to its surroundings, and hence its stochastic trajectory as quantified by its scalar time-dependent displacement, $r(t)$, and is given by [43,49]

$$\beta F_{dyn}(r) = -3 \ln\left(\frac{r}{d}\right) - \frac{\rho}{2\pi^2} \int_0^\infty dk \, \frac{k^2 C(k)^2 S(k)}{(1+S^{-1}(k))} e^{-\frac{k^2 r^2}{6}(1+S^{-1}(k))}. \quad (10)$$

For hard spheres (diameter $d$) calculation of the dynamic free energy requires only $S(k)$. As illustrated in the inset of Fig.5B, the dynamic free energy has a minimum at $r = r_L$ and a maximum at $r = r_B$ with a barrier height of $F_B$ beyond the NMCT crossover ($\phi > \phi_c \sim 0.43$ for hard spheres, for example). For our present purposes, the only essential aspects are the dynamic localization length and the maximum restoring force, $f_{max}$, that occurs at $r = r^*$. The main frame of Fig.5B illustrates the evolution of the key features of the dynamic free energy with attraction strength. The germane trends are: (i) a decrease of the localization length, and hence increase of elastic shear modulus $G'$, with growing attraction (or concentration), and (ii) an increase of the maximum restoring force with attraction (or concentration). These trends reflect the tendency of microgels to cluster and form physical bonds due to a short-range attractive interaction, the dynamical consequences of which reinforce the steric caging effect.

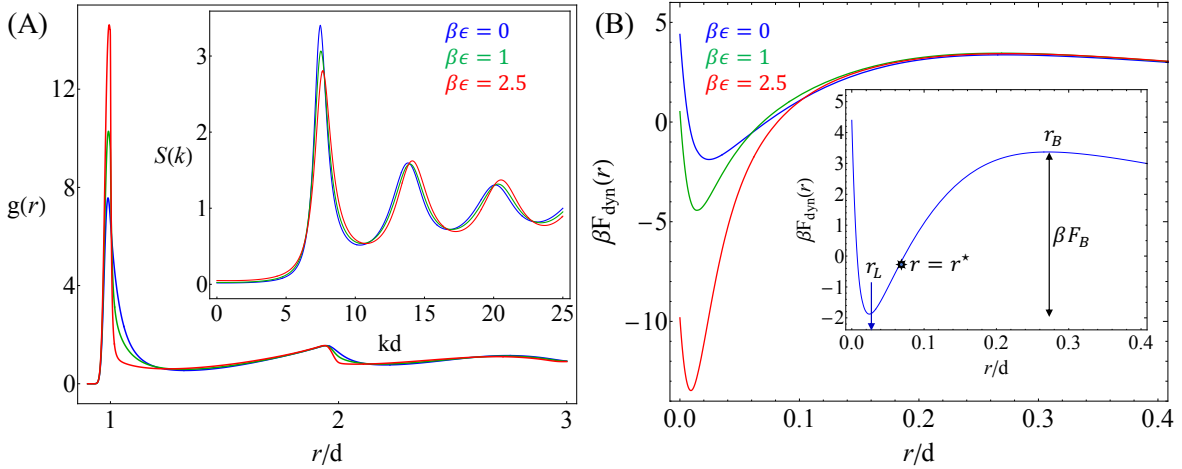

Figure 5 (A) Equilibrium pair correlation function for a microgel suspension of volume fraction ϕ=0.60, and attraction range of a=0.01d. (Inset) static structure factor for the same system parameters at three strengths of attraction. (B) Examples of the dynamic free energy for $\phi = 0.60$, $a = 0.01d$ and strengths of attraction $0k_BT$ (pure Hertzian repulsive case, blue), $1.00k_BT$ (green) and $2.5k_BT$ (red). (Inset) shows the important length scales associated with dynamic free energy and the local barrier.

The above theory of equilibrium dynamics can be extended to treat the effect of external deformation which is manifested as a microscopic force altering the dynamic free energy and single particle trajectories [9,45-46] as $\beta F_{dyn}(r,\sigma) = \beta F_{dyn}(r,0) - A\sigma r$, where $A$ relates the microscopic force to the macroscopic stress ($\sigma$) and $A = \frac{\beta \pi}{6} \phi^{-\frac{2}{3}} d^2$, per prior work [9]. While different choices of the force transduction factor $A$ affect results quantitatively, it does not change the physical picture of external force or stress assisted 'liquification' of the dynamic free energy or any qualitative trends predicted by the theory [45]. With increasing stress, the barrier height decreases as a continuous function of increasing deformation [45]. The critical stress where the localized state (minimum of the dynamic free energy) and barrier simultaneously disappear is termed the "absolute yield stress" ($\sigma_y^{abs}$). It is a simple metric of a mechanically driven solid-to-liquid transition, and equals the theoretically predicted maximum force, $f_{max}$, of the quiescent dynamic free energy divided by cross-sectional area $A$. A simple estimate of an absolute yield strain ($\gamma_y$) can be similarly defined as $\gamma_y = \sigma_y^{abs}/G'(\sigma = 0)$ [9], where $\gamma$ is shear strain [45]. All structural correlations ($S(k)$) are taken to be unaltered at their equilibrium values, a simplification discussed in great detail in prior papers [9,45-46].

## IV. Theoretical Results and Comparison with Experiments

### A. Low Temperature

We now calculate the storage modulus $G'$ for all concentrations in the low temperature regime where no attractions are present. Since the microgels shrink with heating, their effective volume fraction decreases, and hence $G'$ is predicted to decrease, as physically expected (not shown). Given the complex de-swelling behavior and size change of microgels at different concentrations, we focus only on the most dramatic effect: the ratio of the storage modulus at low temperature (10°C) to its value essentially at the LCST temperature (32°C).

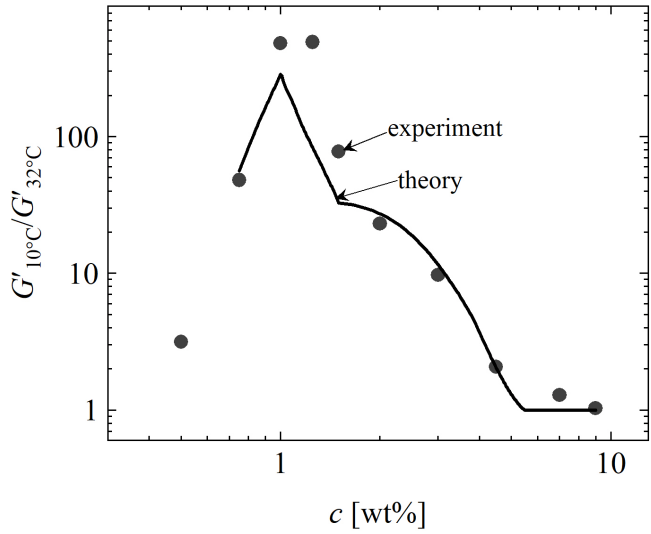

Figure 6: Comparison of the relative change of the shear elastic modulus at low temperatures compared to its value at the LCST. The theoretical result (curve, not a fit) captures the trends observed in the experiments (points, taken from Fig. 3A).

Figure 6 shows experimental and theoretical results for the ratio of the storage modulus at the low temperature of 10°C to its high temperature value at the LCST where $G'(T)$ exhibits a minimum (see Fig. 3A). The data show a dramatic non-monotonic evolution of the ratio with concentration over 3 orders of magnitude. Qualitatively, the theory and experiment show the same trends, even though the precise numbers unsurprisingly modestly differ. At low concentrations, since the effective volume fraction at high temperature is relatively low, the systems are very close to the NMCT crossover where the solid-like behavior disappears (no predicted localized state). At higher concentrations, c > 5.5 wt%, since there is no temperature dependent de-swelling, the ratio is trivially unity, as both size and volume fraction become temperature independent. An interesting observation is that both theoretical predictions and experimental observations find the ratio $\frac{G'(10°C)}{G'(32°C)}$ is strongly non-monotonic with concentration *and* achieves a maximum around the onset of the soft jamming crossover. This dramatic behavior is presumably due to a major change in volume fraction between the high and low $T$ regimes, which maximizes around the soft jamming crossover as shown in Fig. 2B. Considering the *no-fit* nature of the comparison between theory and experiment, and our simple modeling of how microgel size (and hence suspension effective volume fraction) evolves with temperature and concentration, we consider the level of agreement rather remarkable. We believe it provides strong support for the proposed physical idea that there are highly nontrivial and rich consequences of modest microgel de-swelling as a precursor of the LCST transition.

### B. High Temperature

As discussed above and from Fig. 3A, concentration dependent de-swelling at low $T$ results in a microgel of size comparable to its high $T$ collapsed state value. In other words, within the studied temperature and concentration ranges, concentration-dependent de-swelling and temperature-dependent de-swelling have similar effects on the particle diameter or, $d(c_{min}, T_{max}) \equiv d(c_{max}, T_{min}) = d_0$). From Fig. 3A we also note that for low concentrations up to 1.25 $wt\%$, the storage modulus for $T > T_{LCST}$ does not show a gradual increase, presumably due to the very low effective volume fraction of the samples as we estimated in Fig. 2B. Starting from $c = 1.5\ wt\%$ and extending to the highest concentration studied of $9wt\%$, we find a gradual increase of the storage modulus as one heats above the LCST. Due to our inability to a priori predict quantitatively how to switch from repulsive to attractive interactions as a continuous function of temperature, we focus on understanding the concentration dependence of the storage modulus only at highest $T = 40°C$ where inter-microgel attractions are strongest and the consequences of physical bond formation is most dramatic. We note that within the studied range of concentration, $G'(40°C)$ grows enormously by almost ~ 5 orders of magnitude. Fig. 7A shows the experimental data at 40°C , along with our theoretical attempt to understand it.

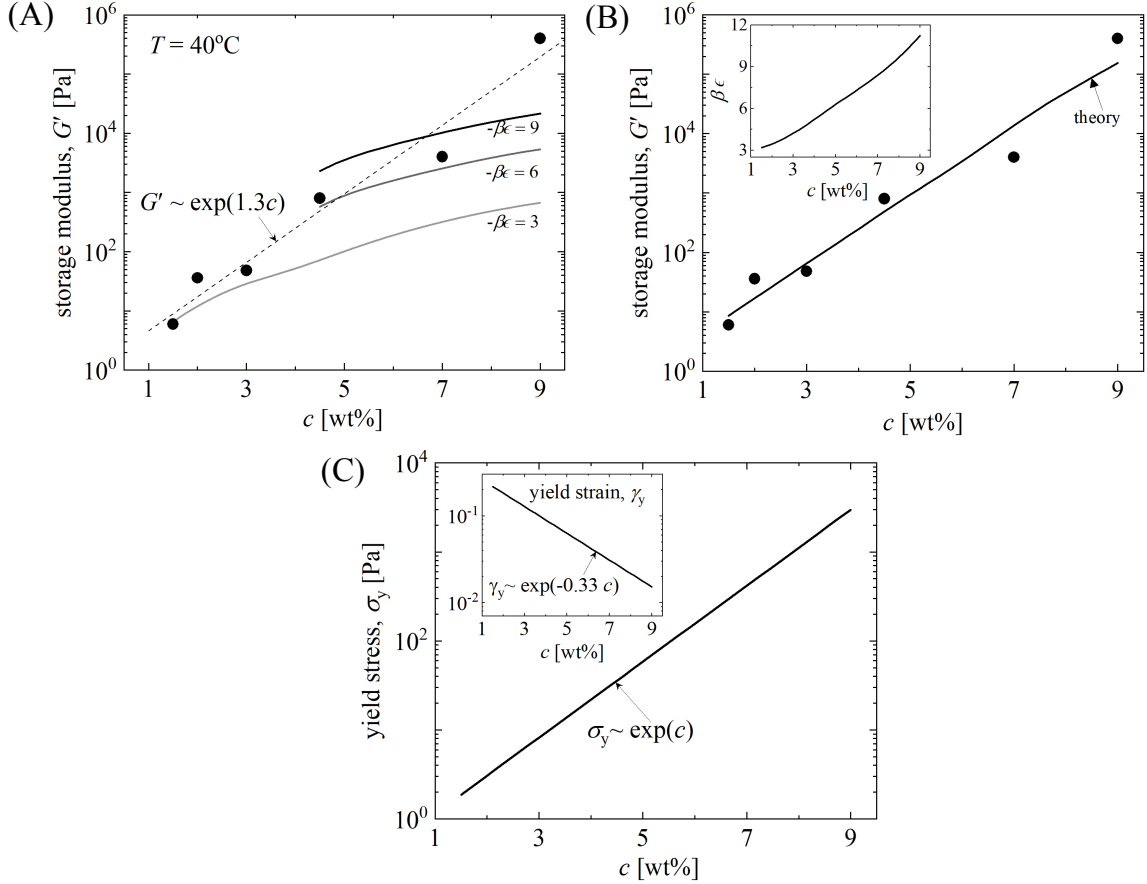

Figure 7 (A) Storage modulus at 40°C. Black dots are experimental points from Fig. 3A, and the black dashed line is an exponential fit to the data. Theoretical calculations (gray curves) are performed at fixed strength of attractions of $-\beta\epsilon = 3, 6, 9$ and fixed range of interaction $a = 0.01d$. (B) Same experimental data compared to theoretical calculations with a concentration-dependent strength of attraction (inset) that reproduces the experimental trend. (C) Theoretical prediction of the absolute yield stress and yield strain (inset) as a function of concentration at $T = 40°C$ where attraction strength depends on concentration per Fig. 7(B) inset.

Given the discrete and limited nature of the experimental storage modulus data (black points) and the log-linear form of the plot, we consider a least square fit to an exponential form to be the best representation of concentration dependence of the $G'$ data, and find $G' \approx 1.21 e^{1.33c}$, where $c$ is the concentration in $wt\%$ and $G'$ is in units of $Pa$. Our theoretical analysis aims to understand this modulus growth per the black dashed line in Figure 7. To convert the dimensionless theoretical $G'$ into an absolute modulus in Pascals, we (i) employ a numerical prefactor of 0.1 in Eq(5) which is known to compensate for the quantitative over prediction of $G'$ by NMCT (the same approach used to analyze low temperature concentration dependence in our previous work [9]), and (ii) convert $\frac{k_B T}{d^3}$ to $Pa$ units using $\frac{k_B T}{(250nm)^3} \approx 0.256 Pa$ where 250nm is the high T collapsed size of the microgels independent of concentration. The Hertzian model parameter $E$ in Eq. (9) is taken to be unchanged, i.e. $E = 30{,}000$ as mentioned earlier.

Theoretical results are presented in Fig. 7A for constant attraction strengths of $-\beta\epsilon = 3$, $6, 9$ at fixed spatial range $= 0.01d$. They show a much weaker growth of the elastic (storage) modulus as a function of concentration than experimentally observed. We believe this reflects a concentration dependence of the attractive inter-particle hydrophobic interaction. This seems physically natural since the effective pair potential is a water mediated potential-of-mean force between microgels, and hence is expected to depend on microgel concentration. A priori, one might expect the effective hydrophobic attraction increases with microgel loading. To empirically account for this many body effect, we calibrate the theoretical model to reproduce the experimental $G'$ data by allowing the strength of attraction, $\epsilon$ in Eq.5, to increase with microgel concentration. We have also studied other attraction ranges (not shown) and find (as expected) that any reasonable variation does not change our results significantly. An estimate of volume fractions at the highest $T = 40°C$ is given in Fig. 2B where the collapsed size is taken to be $2R_{collapsed} = 250nm$ and Eq (10) is used to determine temperature dependent volume fraction. We note that, unlike the low $T$ repulsive microgel case, at high temperature the microgel volume fraction grows almost linearly with concentration.

Fig. 7B plots the theoretical linear elastic shear modulus *fits* to the experimental observations which allow the attraction strength to grow with increasing concentration. The inset shows the deduced concentration dependence of the latter, which grows from $3k_BT$ to $11k_BT$ over the concentration range of 1.5wt% to 9wt%. The black solid line corresponds to theoretical calculations for a fixed attraction range of $a = 0.01d$. We can calculate the structural alpha relaxation time from the dynamic free energy as the time required for barrier hopping via Kramers formula [9,38], as discussed in detail in our prior study of dense repulsive microgel suspensions. For all the concentrations considered here, we have verified that the theoretical alpha relaxation time (an indicator of mean stress relaxation time) is (much) larger than 100 seconds, and hence there is no flow on the experimental time scales, consistent with our present (and prior) viscoelastic measurements.

In an attempt to qualitatively rationalize the concentration dependence of the strength of attraction, we consider the nature of hydrophobic interaction for microgels in poor solvents. The interaction strength per unit area between two identical hydrophobic *flat* surfaces (relevant since our microgels are large) at a surface-to-surface distance $D$ is given by an exponential decay law [34], $W_h(D) = 2\alpha H_y e^{-\frac{D}{D_h}}$, where $W_h$ is the hydrophobic energy per unit area, $\alpha$ is the interfacial free energy of hydrocarbons in water, $H_y$ is the hydrophobicity parameter (equals unity for a perfectly hydrophobic surface, taken to be the case here at T=40°C), and $D_h$ is the short (nm-scale) hydrophobic decay length. The interfacial energy of a hydrocarbon in water is ~50kJ/mol/nm² at 298K [41]. As concentration increases, the surface area of contact between two microgels with fluctuating surfaces is expected to change. If this change is in the direction of a larger contact area, then the total interaction energy (calculated as $W_h \times A_c$, where $A_c$ is the area of contact) would increase. This is one crude physical scenario for our theoretical need to allow the effective attraction between two microgels to grow with concentration in order to understand $G'(c)$. A theoretical estimate for the area of hydrophobic interaction considering $D = D_h = 1 nm$ yields the contact area for Hertzian particles to be $0.43 - 1.57\ nm^2$ in order to generate the desired strength of attraction between particles. This corresponds to the relevant length of order one nm, a seemingly sensible value for this water mediated effect. We note that this discussion assumes the same value for interfacial energy at all concentrations, a simplification that may not be true for the relevant (unsolved) problem of a dense strongly interacting suspension of sticky microgels with fluctuating interfaces. Another scenario is the interfacial energy associated with fuzzy surface microgel particles could grow with increasing suspension concentration (packing fraction) due to osmotic forces and enhanced local interpenetration, which would be another possible mechanism for a growing effective attraction with suspension concentration.

Finally, we can theoretically predict the absolute yield stress and strain which signals the complete destruction of localizing dynamical constraints and a mechanically induced transition of the microgel suspension to a fluid state following the same approach as discussed in detail in refs.43 and 44. Results as a function of concentration (strength of attraction increases with concentration per Fig. 7B) are shown in Fig. 7C. The growth of the absolute yield stress is exponential ($\sigma \sim \exp(c)$), the same form as found for the elastic shear modulus ($G' \sim \exp(1.33c)$), with only a small difference in the numerical prefactor in the exponential. The inset shows the corresponding simple estimates of the absolute yield strain defined as $\sigma_{abs}^y/G'(\sigma = 0)$, which is predicted to weakly decrease with concentration. Since a very short-range attraction is considered between particles, we expect the material to be 'brittle', and thus the theoretical prediction of small yield strains that decrease as the suspensions stiffen via a concentration increase seems qualitatively consistent.

Since our experimental measurements focused only on the linear modulus at high $T$, we do not have any data to compare our theoretical results for yield stresses and strains against. But they do serve as testable predictions for future measurements. For the purpose of this work, predictive plots for *dynamic* yield stresses and strains are not shown. However, as discussed for purely repulsive microgel systems at low $T$ in our prior work [9], theoretical results based on the absolute and dynamic yield criteria do follow the same qualitative trends.

### V.     Conclusions and Summary

We have studied using an integrated experimental and theoretical approach the dynamical and rheological response of concentrated suspensions of thermoresponsive pNIPAM microgel particles over a wide range of concentrations and temperatures. The particles are slightly charged and self-crosslinked, distinct from ionic microgels that have been utilized in prior studies [13, 22], a feature that we have demonstrated in our prior work [9] that leads to several key distinct features in the rheology. Temperature change is shown to have two key effects: (1) the size of individual microgel particles decreases with temperature, and hence the effective volume fraction of suspensions decreases with temperature in addition to the concentration-dependent steric effects analyzed in earlier work [9], and (2) the interparticle interaction changes from purely repulsive to repulsive plus attractive as the temperature is raised above the LCST. While a few previous studies have experimentally studied the response of such thermoresponsive suspensions in highly packed conditions [8, 18, 57], our work provides new insights into the coupled concentration- and temperature-dependent dynamics and mechanics of these systems over a very wide range of conditions.

The concentration regime that we have explored spans the modest viscosity liquid, glassy fluid, and "soft jammed" regimes at low temperatures (below the LCST) where the microgels interact via repulsive interactions. Systematic study of the effect of temperature on the linear and nonlinear rheological properties of the suspensions over the entire concentration regime reveals that the coupled effects of (1) and (2) results in a rich and diverse dynamical behavior, which to the best of our knowledge has not been explored in previous studies. To understand the underlying physics, we have employed a suite of existing statistical mechanical theories, along with the knowledge of single microgel particle size, and simple approximations for interparticle pair-potential, to make quantitative predictions for structural and rheological properties. A temperature-independent Hertzian potential function was employed to model the repulsive-glassy behavior of the suspensions below the LCST, while an additional exponentially decaying short range attractive potential was added to address the above LCST regime. The theoretical predictions provide a consistent understanding of our observations for the linear elastic shear modulus and yield stress over the entire concentration regime. Testable predictions for several other structural and rheological properties are also made.

We believe the ideas developed in the present work are applicable to a diverse range of suspensions, both repulsive and attractive. Our approach also suggests guidelines for designing the key rheological properties of such suspensions from the elementary parameters such as single particle stiffness, polymer-solvent interactions, and interparticle interactions, all of which in principle can be controlled independently.

**Acknowledgement**

This research is supported by the U.S. Department of Energy, Office of Basic Energy Sciences, Division of Materials Sciences and Engineering under Award Nos. DE-FG02-07ER46471 and DE-SC0020858, through the Materials Research Laboratory at the University of Illinois at Urbana-Champaign. RHE thanks Anton Paar for providing the MCR702 rheometer which was used for some of the rheology experiments.

# Supplementary Information

# Linear and Nonlinear Viscoelasticity of Concentrated Thermoresponsive Microgel Suspensions

Gaurav Chaudhary[1,4,7+], Ashesh Ghosh[2,4,8+], Jin Gu Kang[3,4], Paul V. Braun[1-5], Randy H. Ewoldt[1,4,5*] and Kenneth S. Schweizer[2-6*]

[1]Department of Mechanical Science and Engineering, University of Illinois at Urbana-Champaign, Urbana, IL, 61801
[2]Department of Chemistry, University of Illinois at Urbana-Champaign, Urbana, IL, 61801
[3] Department of Materials Science and Engineering, University of Illinois at Urbana-Champaign, Urbana, IL, 61801
[4] Materials Research Laboratory, University of Illinois at Urbana-Champaign, Urbana, IL, 61801, US
[5] Beckman Institute for Advanced Science and Technology, University of Illinois at Urbana-Champaign, Urbana, IL, 61801, USA
[6] Department of Chemical & Biomolecular Engineering, University of Illinois at Urbana-Champaign, Urbana, IL, 61801, USA
[7] Present Address: School of Engineering and Applied Sciences, Harvard University, Cambridge, MA 02138, USA
[8] Present Address: Department of Chemical Engineering, Stanford University, CA 94305, USA

*ewoldt@illinois.edu
*kschweiz@illinois.edu

+ these authors contributed equally to this work

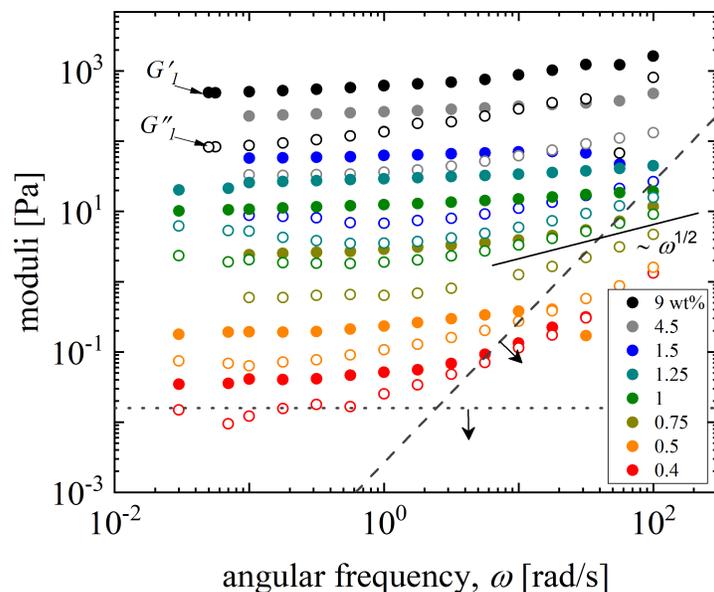

**Figure S1** Frequency dependence of storage and loss moduli measured at strain a amplitude of 1% (linear response regime) at $T = 10$ °C. Suspensions at $c > 0.4$ wt% do not flow on the longest probed time scales (~100 s) (figure adapted from [9]). Experimental limits are shown by the dashed horizontal line (minimum torque limit) and the dashed line (instrument inertia limit).

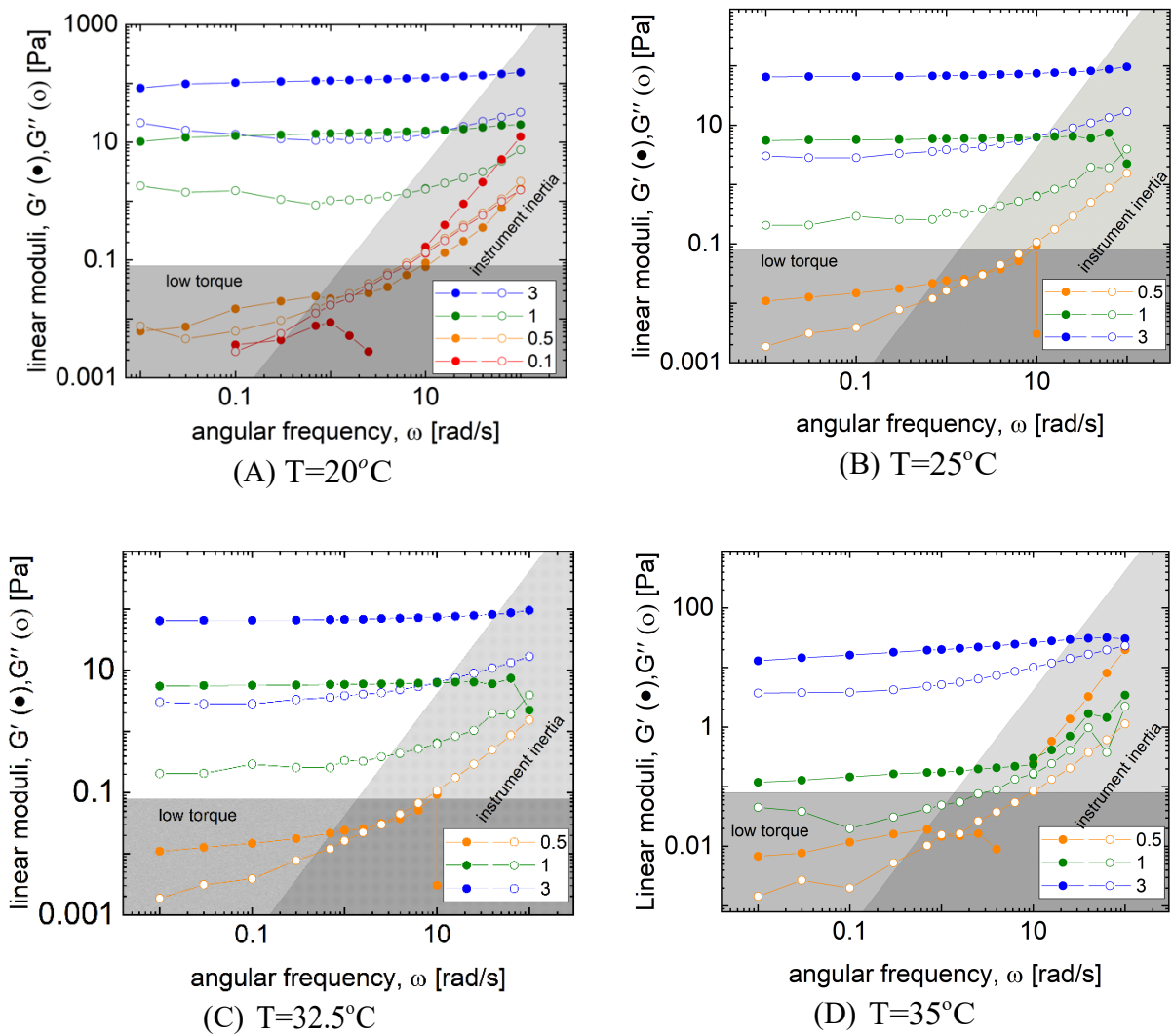

**Figure S2** Frequency dependence of viscoelastic moduli at a strain amplitude of 1% at different temperatures as indicated. Legends show the sample concentrations in wt%. The measurable response indicates a nearly frequency independent behavior. Minimum torque limit and instrument inertia limit are indicated in the gray region.

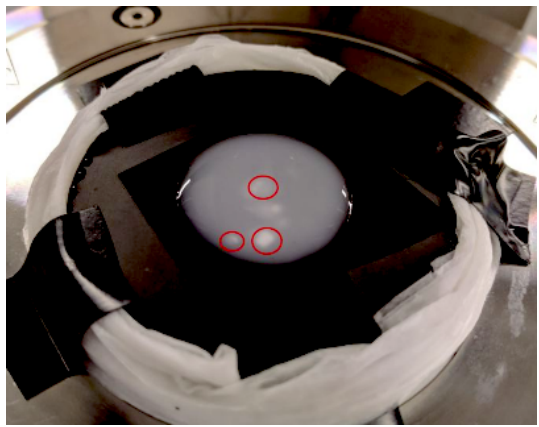

**Figure S3** Visualization of sample. Above the LCST, microgels aggregate due to attractive interactions.

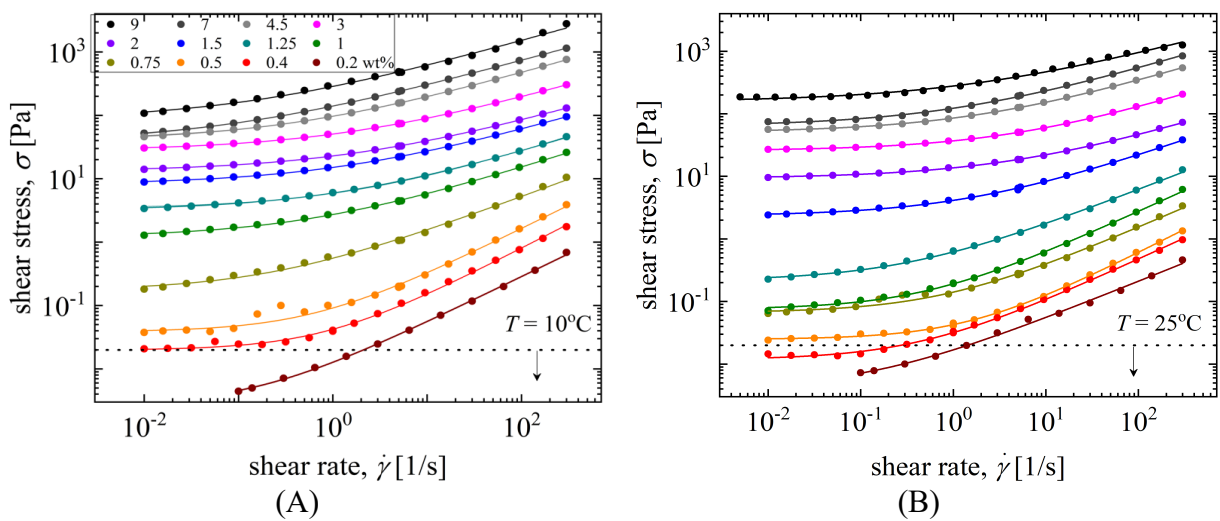

**Figure S4** Steady state shear flow curves for various indicated suspension concentrations at (A) 10°C and (B) 25 °C. All the suspensions above $c$ = 0.4 wt% show an apparent yield stress fluid-like response, with a plateau value of stress at low shear rate. The solid lines show the Herschel-Bulkley variance-weighted fits to the experimental data.

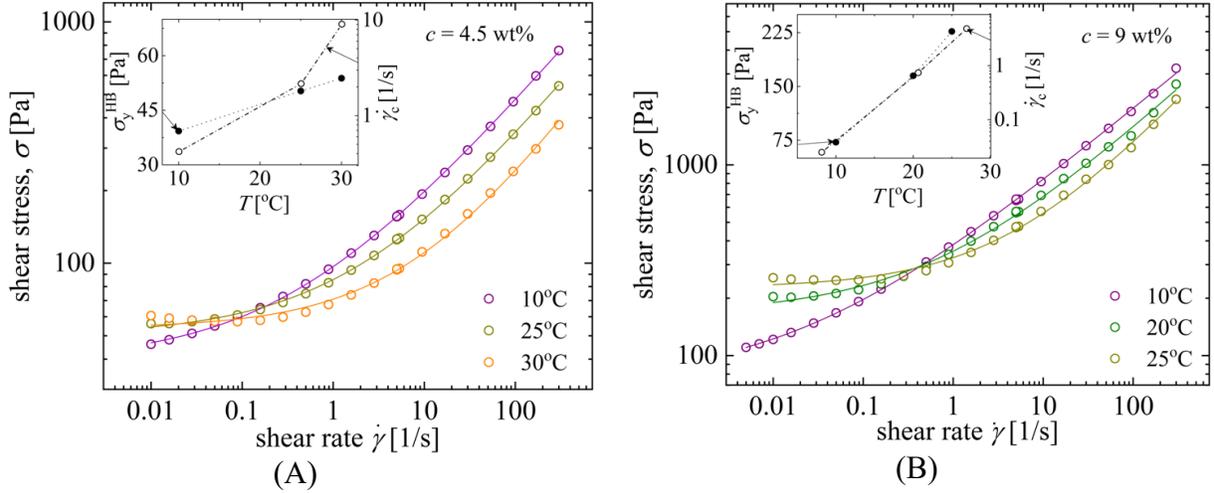

**Figure S5** Temperature dependent flow curves for suspension concentrations (A) 4.5 wt% and (B) 9 wt%. Nonmonotonic behavior is observed at low shear rates. Insets show selected Herschel-Bulkley model fit parameter values.

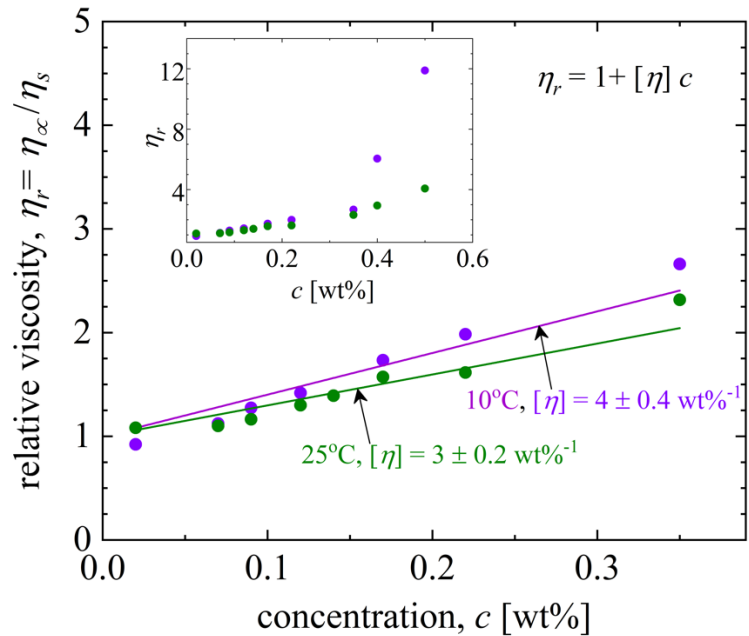

**Figure S6** Relative viscosity, $\eta_r = \eta_\infty / \eta_s$, for low concentration suspensions using an "infinite" shear rate assumed equal to the highest shear rate used, 300 s$^{-1}$. The data agrees well with the Einstein equation ($\eta_r = 1 + 2.5\phi$). For dilute suspensions ($c \to 0$), the effective volume fraction can be related to the mass fraction (wt%) using, $2.5 = [\eta]c$, where $[\eta]$ is the intrinsic viscosity and $\eta_s$ is the pure solvent viscosity (= 0.0091 Pa s). We obtain variance weighted best fit values of $[\eta] = 4 \pm 0.4 wt\%^{-1}$ and $[\eta] = 3 \pm 0.2 wt\%^{-1}$ at $10^oC$ and $25^oC$, respectively.